
\input harvmac.tex
\input amssym.def
\input amssym
\def\capt#1{\narrower{
\baselineskip=14pt plus 1pt minus 1pt #1}}


\lref\sm{Smirnov, F.A.: Form-factors in 
Completely Integrable Models
of Quantum Field Theory, World Scientific, Singapore, 1992}

\lref\Grun{Gruner, G.: Density Waves in Solids, Addison-Wesley, 1994}
 
\lref\Fab{Essler, F.H.L. and Tsevelik, A.M.:
Weakly coupled one-dimensional Mott insulator.
Preprint, 2001  (cond-mat/0108382) }

\lref\Belavin{Belavin, A.A.: Exact solution of
the two-dimensional model with asymptotic freedom. 
Phys. Lett.  {\bf B87}, pp. 117-121 (1979)} 

\lref\dotse{Dotsenko, V., Picco M. and Pujoi, P.:
Renormalization group calculation of correlation
functions for the 2D random bound Ising and Potts models.
Nucl. Phys. {\bf B455}, pp. 701-723 (1995) }

\lref\gr{Guida, R. and Magnoli, N.: Tricritical Ising model near
criticality. Int. J. Mod. Phys. {\bf A13}, pp. 1145-1158 (1998) }
 
\lref\Bour{Bourbonnais, C. and Jerome, D: The normal phase
of quasi-one-dimensional organic superconductors. 
In ``Advances in Synthetic
Metals, Twenty years of Progress in Science and
Technology'', ed. by Bernier P., Lefrant, S. and Bidan, E. 
Elsevier, New York, 1999}

\lref\Jin{Zinn-Justin, J.: Quantum Field Theory and Critical
Phenomena, Clarendon Press, Oxford, 1996 (third ed.)}

\lref\drew{
Babujian, H., Fring, A.,  Karowski, M. and  Zapletal, A.:
Exact form-factors 
in integrable quantum field theories: the sine-Gordon model.
Nucl. Phys. {\bf B538}, pp. 535-586 (1999)}

\lref\delf{
Delfino, G.: Off critical correlations in the Ashkin-Teller model.
Phys. Lett. {\bf B450}, pp. 196-201 (1999)   }

\lref\ZamolAl{ Zamolodchikov, Al.B.:
Two-point correlation function
in scaling Lee-Yang model.
Nucl. Phys. {\bf B348}, pp. 619-641 (1991)}

\lref\coloda{
Acerbi C., Mussardo G., and Valleriani, A.:
Form-factors and 
correlation functions of the stress - energy tensor in massive
deformation 
of the minimal models $E_{(N)}^{(1)}\times E_{(N)}^{(1)}/E_{(N)}^{(2)}$.
Int. J. Mod. Phys. {\bf A11}, pp. 5327-5364 (1996)}

\lref\colodaa{Balog, J. and  Niedermaier, M.:
Off-shell dynamics of the 
O(3) nonlinear sigma model beyond Monte Carlo and perturbation theory.
Nucl. Phys. {\bf B500}, pp. 421-461 (1997) }

\lref\Korep{Korepin, V.E. and Essler, F.H.L.:
$SU(2)\otimes SU(2)$ invariant scattering matrix of the Hubbard model.
Nucl. Phys. {\bf B426}, pp. 505-533 (1994)} 

\lref\Woyn{ 
Woynarovich F.: Excitations with 
complex wave numbers in a Hubbard chain. 1. States with one pair of
complex wave numbers. J. Phys. {\bf C}16, pp. 5293 (1983)\semi
Excitations with 
complex wave numbers in a Hubbard chain. 2. States with 
several pairs
of complex wave numbers. 
J. Phys. {\bf C}16, pp. 6593 (1983)}

\lref\lutem{Emery, V.J., Luther, A.  and Peschel, I.:
Solution of the one-dimensional electron gas on a lattice.
Phys. Rev. {\bf B13}, pp.1272-1276 (1976)}
 
\lref\fil{Filev, V.M.: 
Spectrum of two-dimensional relativistic model.
Teor. i Mat. Fiz. {\bf 33}, pp. 119-124 (1977)}

\lref\Mel{Melzer, E.:
On the scaling limit of the 1-d Hubbard model at half filling.
Nucl. Phys. {\bf B443}  [FS], pp. 553-564 (1995)}

\lref\Woynarovich{Woynarovich, F. and  Forgacs, P.:
Scaling limit of the one-dimensional attractive Hubbard model: 
The half-filled band case.
Nucl. Phys. {\bf B498}, pp.65-603 (1997)}
 
\lref\AliGracey{
Ali, D.B. and Gracey, J.A.:
Four loop wave function renormalization in the non-abelian Thirring model.
Nucl. Phys. {\bf B605}, pp. 337-364 (2001)}

\lref\BennettGracey{
Bennett, J.F. and Gracey, J.A.:
Three-loop renormalization of the $SU(N_c)$ non-abelian Thirring model.
Nucl. Phys. {\bf B563}, pp. 390-436 (1999)
}

\lref\QCD{Stevenson, P.M.: Optimized perturbation theory. Phys. Rev. {
\bf D23}, pp. 2916-2943 (1981)}

\lref\colm{Coleman, S.: The quantum sine-Gordon
equation as the Massive Thirring model. Phys. Rev.{\bf D11},
pp. 2088-2097 (1975)}

\lref\Controzzi{
Controzzi, D., Essler, F.H.L. and Tsevelik, A.M.:
Optical conductivity of one-dimensional Mott insulators,
{\sl preprint cond-mat/0005349}
}

\lref\CGHJK{Corless, R.M., 
Gonnet, G.H., Hare, D.E.G., Jeffrey, D.J. and Knuth, D.E.: On the Lambert
W function. Adv. Comput. Math. {\bf 5}, no.4, pp. 329-359 (1996)\semi
Jeffrey, D.J., Hare, D.E.G. and Corless, R.M.:
Unwinding the branches of the Lambert W
function. Math. Sci. {\bf 21}, pp. 1-7 (1996)
}

\lref\gog{Gogolin, A.O., Nersesyan, A.A. and Tsvelik, A.M.:
Bosonization and Strongly Correlated Systems.
Cambridge University Press, 1999}

\lref\Mandelstam{Mandelstam, S.:
Soliton operators for the quantized sine-Gordon equation.
Phys.Rev. {\bf D11}, pp. 3026-3030 (1975)}

\lref\lik{Lukyanov, S.: Free field 
representation for massive integrable models.
Commun. Math. Phys. {\bf 167}, No.1, pp. 183-226  (1995) \semi
Form-factors of exponential fields
in the sine-Gordon model.
Mod. Phys. Lett. {\bf A12}, pp. 2543-2550 (1997)}

\lref\LZtopff{
Lukyanov, S. and Zamolodchikov, A.:
Form factors of soliton-creating operators in the sine-Gordon model.
Nucl. Phys. {\bf B607} [FS] pp. 437-455 (2001)}

\lref\Mathur{
Mahur, S.D.:
Quantum Kac-Moody symmetry in integrable field theories.
Nucl. Phys. {\bf B369}, pp. 433-460 (1992)
}

\lref\Woynarovich{Woynarovich, F. and  Forgacs, P.:
Scaling limit of the one-dimensional attractive Hubbard model: 
The half-filled band case.
Nucl. Phys. {\bf B498}, pp.65-603 (1997)}

\lref\Zms{
Zamolodchikov, Al.B.:
Mass scale in the sine-Gordon model and its reduction.
Int. J. Mod. Phys. {\bf A10}, pp. 1125-1150 (1995)}

\lref\Zmss{
Zamolodchikov, Al.B.: unpublished
}

\lref\exper{Hwu, Y. et al.: Photoemission near the Fermi energy
in one dimension. Phys. Rev. {\bf B46}, pp. 13624-13626 (1992)\semi
Zwick, F. et al.: Absence of quasiparticles in
the photoemission spectra of
quasi-one dimensional Berchgaard salts.
Phys. Rev. Lett. {\bf 79}, pp. 3982-3985 (1997)}

\lref\Andrei{Andrei, N. and Lowenstein, J.H.:
Diagonalization of the Chiral-Invariant Gross-Neveu Hamiltonian.
Phys. Rev. Lett. {\bf 43}, pp. 1698-1701 (1979)}

\Title{\vbox{\baselineskip12pt\hbox{RUNHETC-2002-7}}}
{\vbox{\centerline{}
\centerline{ Fermion Schwinger's function}
\centerline{  for the  $SU(2)$-Thirring model } }}
\centerline{}
\centerline{Benjamin Doyon$^1$ and  Sergei Lukyanov$^{1,2}$ }
\centerline{}
\centerline{$^1$Department of Physics and Astronomy,
Rutgers University}
\centerline{ Piscataway,
NJ 08855-0849, USA}
\centerline{}
\centerline{$^2$L.D. Landau Institute for Theoretical Physics}
\centerline{Kosygina 2, Moscow, Russia}
\centerline{}
\centerline{}

\bigskip
\centerline{\bf{Abstract}}

\bigskip

We study the Euclidean two-point function of Fermi fields
in the $SU(2)$-Thirring model on the whole
distance (energy) scale. We perform perturbative and
renormalization group analyses to obtain the short-distance
asymptotics, and numerically evaluate the
long-distance behavior by using the form factor
expansion. Our results illustrate the use
of bosonization and conformal perturbation theory
in the renormalization group analysis of a fermionic
theory, and numerically confirm the validity
of the form factor expansion
in the case of the $SU(2)$-Thirring model.

\bigskip
\Date{May, 02}

\eject

\newsec{Introduction}

Correlation functions in 2D integrable 
models have attracted
much attention from experts in Quantum Field Theory (QFT).
The possibility of their
exploration on the
whole length (energy) scale is of great importance.
It gives a rare  opportunity to
probe non-perturbatively  general principles
of QFT. There is also   a  pragmatic  reason for this interest.
The past two decades have witnessed experimental
work to identify and study quasi one-dimensional
systems (for a review, see\ \refs{\Grun,\Bour}). 
There were collective efforts of many physicists to
apply integrable QFT   to describe
such physical systems\ \gog. 
For this purpose, a non-perturbative
treatment of the
correlation functions in 
integrable models seems to be
valuable. It is worth noting that
in recent years,
angle resolved photoemission spectroscopy
has matured into a powerful experimental method for probing
the  electronic Green's functions in quasi one-dimensional
systems\ \exper. Hence  two-point fermion  correlators  in integrable   
theories deserve special consideration.

In this paper we are studying
Schwinger's  function (Green's function in the  Euclidean
region)
$$\big\langle\, \Psi_{\sigma}(x)\, {\bar \Psi}_{\sigma'}(0)\, 
\big\rangle$$
in the  $SU(2)$-Thirring model,
which is described by the   Euclidean 
action\foot{The definition of the coupling constant $g$
is not conventional, but convenient for our purposes.}
\eqn\thirringi{{\cal A}=
\int\, d^2x\ \bigg\{\, \sum_{\sigma=\uparrow,\downarrow}\, 
{\bar\Psi}_{\sigma}
\gamma^{\mu}\partial_{\mu}\Psi_{\sigma} + {{\pi g}\over
8}\
\big({\bar\Psi}\gamma_{\mu}{\vec \tau}\Psi\big)^2\, \bigg\}\ .}
Here $\Psi_{\sigma}$ is a doublet of Dirac
Fermi fields, and
the Pauli matrices ${\vec \tau} = (\tau^1,
\tau^{2}, \tau^{3})$  act on the ``colour'' indices
$\sigma=\, \uparrow, \downarrow$. 
The QFT\ \thirringi\ possesses a variety of 
interesting properties\ \Jin. For instance,
it is an asymptotically free theory (for $g>0$) with unbroken chiral
symmetry, and its   mass scale $M$
appears through dimensional
transmutation. The model  belongs to the very special class of
field theories which admit an infinite
number of conservation laws preventing particle
production in scattering processes.
The Hamiltonian  of\ \thirringi\  was diagonalized
by the  Bethe Ansatz techniques
in \refs{\Andrei, \Belavin}.

Also, it  is  a  popular model
for the interacting one-dimensional electron gas; 
as is known \refs{\fil,\lutem,
\Korep,
\Mel,\Woynarovich},\ \thirringi\
describes the scaling limit of the half-filled Hubbard chain,
$${\bf H}_{Hub}=-\sum_{j=-\infty}^{+\infty}\bigg\{ \sum_{\sigma =
\uparrow, \downarrow} \big(\, c_{j ,\sigma}^\dagger
c_{j+1, \sigma} +
c_{j+1 ,\sigma}^\dagger
c_{j, \sigma}\, \big)+U \ 
\Big(\, c^{\dagger}_{j,\uparrow}c_{j ,\uparrow}-{1\over 2}
\, \Big)\Big(\, c^{\dagger}_{j,\downarrow}c_{j ,\downarrow}-{1\over 2}
\, \Big)\, \bigg\}\, .$$
where
$\{\, c_{j ,\sigma}^\dagger\, ,c_{j' ,\sigma'}\, \}=
\delta_{\sigma\sigma'}\ \delta_{jj'}. 
$
More precisely, if one sends the coupling constant $U \to +0$, 
the correlation length
$$
R_c = {\pi\over 2\sqrt{U}}\  e^{2\pi \over U}
$$ 
diverges and the correlation functions in the Hubbard chain 
at large lattice separations 
assume certain scaling forms. In particular, 
if\ $|j-j'|\gg 1$, the equal-time fermion correlator
can be written as
\eqn\oip{\langle\, c_{j' ,\sigma'}\, 
c_{j ,\sigma}^\dagger\, \rangle \to \delta_{\sigma'\sigma}\ 
{{\sin\big(\, {\pi\over
2}\, (j'-j)\, \big)}\over{\pi\,  (j'-j)}} \ \  F(\, |j'-j|/R_c\,)\, .}
The scaling function $F$ here is directly related
to the field-theoretic correlation function:
\eqn\piu{\langle\, \Psi_{\sigma'}(x)\, {\bar \Psi}_{\sigma}(0)\, \rangle=
{\delta_{\sigma'\sigma}\over 2\pi}\, {\gamma_{\mu}x^{\mu}\over 
|x|^2}\, F(M|x|)\ .}

Our analysis of the fermion correlator is based, 
on the one hand, on recently proposed
expressions for the
form factors of soliton-creating operators
(or topologically charged fields) in the
sine-Gordon model\ \LZtopff\foot{Without
taking normalization into consideration, some of such form factors were
considered previously in Refs. \refs{\drew,\delf}}, and on the other hand, 
on a conformal perturbative analysis
of two-point correlation functions involving such fields. 
The form factor expressions
can be used to obtain the long-distance behavior 
of these two-point functions,
whereas Conformal Perturbation Theory (CPT) gives their short-distance
expansion\ \ZamolAl.
The interest in some of these topological fields stems from
their r\^ole   in fermionic theories.
For instance,
it is well-known that the sine-Gordon model is equivalent to the
massive Thirring model \colm.
The components
of the Thirring fermion field
are then associated with soliton-creating operators of
topological charge
$\pm1$ and Lorenz spin $\pm{1\over 2}$,
and correlators of these operators in the sine-Gordon model
are  related to fermion correlators in the
massive Thirring model\ \Mandelstam.
More interestingly, the sine-Gordon theory is closely related
to a model which is an integrable deformation
of\ \thirringi\ \Jin. This
``deformed''  (or anisotropic) $SU(2)$-Thirring model
exhibits the
so-called spin-charge separation, which is translated
by its representation in terms of two bosonic theories,
one for the charge part, one for the spin part. The
spin part of the fermion
field corresponds to soliton-creating
operators of topological charge $\pm1$ and Lorenz spin
$\pm{1\over 4}$ in the sine-Gordon model, 
and its charge part is related to similar operators
in a free massless bosonic theory.

Although form factor expansions and CPT are very effective
tools for the study of, respectively, the long-distance and the 
short-distance asymptotics of Schwinger's functions\ 
\refs{\ZamolAl,\coloda,\colodaa}, one usually gets into
trouble when trying to compare both predictions in a region where
they are expected to be accurate enough. 
Indeed, in general, one has the freedom 
of choosing  the overall multiplicative
normalization in the CPT  expansion
as well as in the form factor  expansion,
and there is no systematic  way of relating  both normalizations. 
For the case of the soliton-creating operators, 
the constant relating both normalizations was conjectured
in \LZtopff. It allows one to make  unambiguous numerical  
predictions on
the  correlation functions of soliton-creating fields on the whole
distance scale using the combined CPT and form factor data.
We performed this calculation 
for the case of the $SU(2)$-Thirring fermion.

The paper is organized as follows. 
In Section 2, we recall some standard results concerning
the anisotropic $SU(2)$-Thirring model and its
relation to the sine-Gordon theory.
In Section 3, the short-distance
behavior of correlators of the soliton-creating operators
is examined by means of CPT.
Here we also perform a   Renormalization
Group (RG) resummation  of 
the perturbative expansion in the vicinity of the
Kosterlitz-Thouless point which corresponds  to the  $SU(2)$
limit  of the fermion theory.
In Section 4,  the perturbative calculation  is  adapted to the
momentum space fermion 
Schwinger's function; 
we give the two-point function in the $SU(2)$-Thirring model
to third order in the running coupling. 
This particular result was  recently obtained by standard 
perturbation theory
in the modified Minimal Subtraction (${\overline {\rm MS}}$)  scheme
\AliGracey\ (calculations in\ \AliGracey\  concern, 
in fact, fermion correlators
in a general non-abelian Thirring model). We then compare
the result of Ref.\ \AliGracey\
with ours and explicitly relate our RG scheme to the
${\overline {\rm MS}}$ scheme. In Section 5, 
the long-distance behavior of the  
fermion correlator in the anisotropic
$SU(2)$-Thirring model is analyzed  by  means of the form factors
given in \LZtopff. 
In Section 6, we examine properties  
of  the fermion spectral density
in the $SU(2)$-Thirring model.
The outcome of our calculations is discussed
in Section 7, where we numerically compare the short-distance
behavior of the scaling function $F$\ \oip, \piu\
(from the
RG analysis) with its long-distance behavior (from
its form factor expansion).

Finally, we note that this work  has an essential overlap 
with  Ref. \Fab, where\ \thirringi\ was considered as
a model of one-dimensional Mott insulators.

\newsec{Bosonization of the anisotropic  $SU(2)$-Thirring model}

The $SU(2)$-invariant Thirring model admits an  integrable
generalization such that the underlying $SU(2)$ symmetry is explicitly
broken down to $U(1)\otimes {\Bbb Z}_2$:
\eqn\suthirring{{\cal A}_{DTM} =
\int\, d^2x\ \bigg\{\, \sum_{\sigma=\uparrow,\downarrow}\,
{\bar\Psi}_{\sigma}
\gamma_{\mu}\partial^{\mu}\Psi_{\sigma} + 
{{\pi g_{\parallel}}\over 8}\ J^{3}_{\mu}J^{3}_{\mu}
+{\pi g_{\perp}\over 8}\,
\Big(\, J^{1}_{\mu}J^{1}_{\mu}+J^{2}_{\mu}J^{2}_{\mu}\, \Big)
\,  \bigg\}\, ,}
where
\eqn\currs{
J^{A}_{\mu} =
{\bar\Psi}\gamma_{\mu}\tau^{A}\Psi }
are  vector currents.
The model\ \suthirring\ is
renormalizable, and
its coupling constants $g_{\parallel},\, g_{\perp}$ should be
understood as ``running'' ones. 
In particular, in the RG-invariant domain $g_{\parallel}
\geq |g_{\perp}|$, all RG
trajectories originate from the line $g_{\perp}=0$ of UV stable fixed
points, and\ \suthirring\ indeed defines a quantum field theory\foot
{The Hamiltonians corresponding to
opposite  choices of the
sign of $g_{\perp}$ are unitary equivalent,  so the sign
of this coupling
does not affect the physical
observables.}.
Hence, in this domain (which is the only one that we discuss here),
each RG trajectory is
uniquely characterized by the limiting value
\eqn\wer{\rho={1\over 2}\ \lim_{\ell\to 0}\,g_{\parallel}(\ell)}
of the running coupling
$g_{\parallel}(\ell )$ at extremely 
short distances ($\ell$ stands for the
length scale), i.e. the theory
\ \suthirring\ depends only on the 
dimensionless parameter $\rho$, besides the
mass scale $M$ appearing through dimensional transmutation.

As is well known\ (see e.g. \refs{\gog,\Jin}),
the model\ \suthirring\ can be bosonized in terms of
the sine-Gordon field $\varphi(x)$,
\eqn\ksiy{
{\cal A}_{sG} = \int d^2x\,
\bigg\{\,  {1\over{16\pi}}\,(\partial_{\nu}
\varphi)^2 -
2\mu\,\cos(\beta\varphi)\, \bigg\}\, , }
with the coupling constant $\beta$ in\ \ksiy\ related
to\ $\rho $ \ \wer\ by
\eqn\betag{\beta^2={1\over 1+\rho}\, ,}
and a free massless boson.
Then the mass scale $M$ is  identified
with the   mass  of the sine-Gordon solitons, 
which is related to the parameter
$\mu$ by \Zms 
\eqn\kytre{\mu={\Gamma({1\over 1+\rho})
\over \pi \Gamma({\rho\over 1+\rho})}\
\bigg[M\, {\sqrt{\pi}\Gamma({1\over 2}+{1\over 2\rho})\over
2\, \Gamma({1\over 2\rho})}\bigg]^{2\rho\over 1+\rho}\ .}
The precise operator
relations between\ \suthirring\ and\ \wer\  can be found in \LZtopff.
In particular, for the two-point fermion
correlator, the bosonization implies that
\eqn\terju{\langle\, 
\Psi_{\sigma}(x)\, {\bar \Psi}_{\sigma'}(0)\, \rangle=
{\delta_{\sigma'\sigma}\over 2\pi}\ { \gamma_{\mu}x^{\mu}\over |x|^{
{3\over 2}}}
\  \  \ \ F^{(1)}_{1/4}(r)\ ,}
where we use the notation   $F^{(n)}_{\omega}\ (n=1,\ \omega=1/4)$
for the real function which
depends only on the distance $r=|x|$ (and implicitly on the
the mass scale $M$ and the parameter $\rho$), and which,
in essence,
coincides with the Euclidean correlator
of  nonlocal topologically charged fields
in the  model\ \ksiy:
\eqn\uytaa{\langle\,  {\cal O}_{-\omega\beta}^{n}(x)\,  {\cal
O}_{\omega\beta}^{-n}(0)\,\rangle =
\Big(\, e^{i\pi}\,
{{{\bar {\rm z}}}\over{ {\rm z}}}\, \Big)^{\omega n}\
\  \
F_{\omega}^{(n)} (r)\,,}
where ${\rm z}
= {\rm x}^1+i{\rm x}^2$, ${\bar {\rm z}} = {\rm x}^1 - i{\rm x}^2$.
Again we refer the reader  to the paper \LZtopff\ for the
precise definition of the field
${\cal O}_{a}^n\ (a=\omega \beta)$.
Here we note that  it  carries  an integer topological
charge $n$, a  scale dimension
\eqn\jsytr{d={2 \omega^2\over 1+\rho}+{n^2\over 8}\, (1+\rho)\ ,}
and a Lorentz spin $\omega n$.

\newsec{Short-distance  expansion}

\subsec{Conformal perturbation theory}

We now turn to the analysis of
the short-distance behavior of the correlator\ \uytaa.
In general, one can examine this behavior
via the operator product expansion, for instance:
\eqn\wsiu{
F_{\omega}^{(n)} (r)= {\Bbb C}_{ \bf I}(r)   +
{\Bbb C}_{\cos(\beta\varphi)}(r)\ \langle\,
\cos(\beta\varphi)\, \rangle + \ldots\ .}
The structure functions 
(${\Bbb C}_{ \bf I}(r)$, ${\Bbb C}_{\cos(\beta\varphi)}(r)$, etc.)
admit power series expansions in  $\mu^2$,
 which can
be obtained by using the standard rules of CPT,
whereas the vacuum expectation
values of the associated operators are 
in general non-analytical at $\mu=0$.
In the perturbative treatment, we regard the sine-Gordon
model\ \ksiy\ as
a Gaussian conformal field theory
\eqn\ooiy{
{\cal A}_{Gauss} = \int d^2x\,
 {1\over{16\pi}}\,(\partial_{\nu}
\varphi)^2 }
perturbed by the relevant operator
$\cos(\beta\varphi)$.
Notice that in the limit $\mu\to 0$,
the  nonlocal topologically charged fields ${\cal O}_a^{n}$ can
be expressed in terms of the right and left moving parts of a
free massless field
$\varphi=\varphi_R({\rm z})-
\varphi_L({\bar{\rm z}})$ governed by the action\ \ooiy:
\eqn\oomega{{\cal O}_{a}^{n}\big|_{\mu\to 0} 
\to {\tilde {\cal O}}_{a}^{n}=
\exp\bigg\{\, i\, 
\Big(\, a-{n\over 4\beta}\, \Big)\, \varphi_R({\rm z})-
i\,   \Big(\,  a+{n\over 4\beta}\, \Big)\, \varphi_L({\bar{\rm z}})\,
\bigg\}\ .}
CPT gives the  structure
function ${\Bbb C}_{ {\bf I}}$\ \wsiu\
in the form
\eqn\muyt{\Big(\, e^{i\pi}\,
{{{\bar {\rm z}}}\over{ {\rm z}}}\, \Big)^{\omega n}\ 
{\Bbb C}_{\bf I}(r) =
\Big\langle\,  {\tilde  {\cal O}}_{\omega\beta}^{-n}(x)
{\tilde  {\cal O}}_{-\omega\beta}^n(0)
\exp\Big(2\mu\, \int'd^2y\, \cos(\beta\varphi)\Big)\,
\Big\rangle_{Gauss}\,  ,}
where\ $\langle\, \ldots\, \rangle_{Gauss}$\  is the expectation
value in the Gaussian theory ${\cal A}_{Gauss}$ and
the exponential is understood as a perturbative series in $\mu$.
In the perturbative  series, the integrals will
have {\it power law}
IR divergences which should be thrown away\ \ZamolAl.
Such a regularization
prescription is indicated by the prime near the integral
symbol. In the absence of logarithmic divergences, throwing away
the divergences is
equivalent to treating the integrals as
analytical continuations in the field dimensions\ \ZamolAl.
Considering only the part of $F_{\omega}^{(n)}$ perturbative in $\mu$,
it is a simple matter
to obtain
\eqn\cptres{ F_{\omega}^{(n)}(r) =
r^{-2d}\, \bigg\{1+
J_n(2\omega\beta^2,-2\beta^2)\  \mu^2\, r^{4-4\beta^2} +
O\Big(r^{8-8\beta^2}, r^{2}\Big)\, \bigg\}\ ,}
where
$d$ is given by \jsytr\ and
\eqn\lsuy{\eqalign{J_n(a,c) = &\int' d^2x d^2y\
        x^{a+{n\over 2}}{\bar x}^{a-{n\over 2}}\,
        (1-x)^{-a-{n\over 2}}(1-{\bar x})^{-a+{n\over 2}}\times\cr &
        y^{-a-{n\over 2}}{\bar y}^{-a+{n\over 2}}\,
        (1-y)^{a+{n\over 2}}(1-{\bar y})^{a-{n\over 2}}\,
        |x-y|^{2c}\ .}}
Two comments are in order here. First,
the next omitted term in the short distance expansion\ \cptres\
comes from either  the next term
in the  perturbative series for\ ${\Bbb C}_{\bf I}$\ 
$\big(O(r^{8-8\beta^2})\big)$
or from  the  leading contribution
of  $\cos(\beta\varphi)$\ $\big(O(r^{2})\big)$
in\ \wsiu.
Therefore, the  $\mu^2$ term written in\ \cptres\ is a leading
correction to the scale invariant part of the
correlation function for   ${1\over 2}<\beta^2<1$ only.
Second, 
in writing\  \cptres\  we
specify the overall multiplicative 
normalization of the 
nonlocal topologically charged field ${\cal O}_{\omega\beta}^n$ by
the condition
\eqn\norm{F_{\omega}^{(n)}(r)\to
 r^{-2d}\ \ \ \ \ \ \ {\rm as}\ \ \ {r \to 0}\ .}

The integral\ \lsuy\  can be calculated using,
for instance, techniques illustrated in \Mathur.
The result can be expressed in terms of two
generalized hypergeometric functions at unity:
\eqn\xcsdew{\eqalign{&A(q,c)={}_3F_{2}(-c, -c-1, 1-q;
-c-q, 2; 1)\, \cr
&B(q,c)={}_3F_{2}(q, q+1, c+2;
c+q+2, c+q+3; 1)\ .}}
With $q = a+ n/2$ and ${\bar q}=a-n/2$, we found:
\eqn\fsre{J_n(a,c) = J^{(1)}+J^{(2)}+J^{(3)}+J^{(4)}\ ,}
where
$$\eqalign{
J^{(1)}=&q{\bar q}\ \Gamma(1-q)\Gamma(1-{\bar q})\, \Gamma(1+c+q)
 \Gamma(1+c+{\bar q})\Gamma^2(-1-c)\times \cr &
 \big(\, \cos(\pi (q-{\bar q}))-\cos(\pi c)
\cos(\pi (q+{\bar q}+c))\big)\
A(q,c)A({\bar q},c)\, ,\cr
J^{(2)}=&{\pi^2 q\Gamma(1+c)\Gamma(1+{\bar q}) \Gamma(1+c+ q)
\Gamma(-1-c-{\bar q})\over \Gamma(q)\Gamma(-{\bar q})
\Gamma(3+c+ {\bar q})}\ A(q,c)B({\bar q},c)\, ,\cr
J^{(3)}=& {\pi^2 {\bar q}\Gamma(1+c)\Gamma(1+ q) \Gamma(1+c+{\bar  q})
\Gamma(-1-c- q)\over \Gamma({\bar q})\Gamma(- q)
\Gamma(3+c+  q)}\ B( q,c)A({\bar q},c)\, ,\cr
J^{(4)}=& -{\pi^2 \Gamma^2(1+ q)\Gamma^2(2+c) \Gamma(-1-c-{\bar  q})
\Gamma(-2-c-{\bar  q})\over \Gamma^2(-{\bar q})\Gamma^2(-c)
\Gamma(2+c+  q)\Gamma(3+c+  q)}\ B(q,c)B({\bar q},c)\,  .}
$$
Notice that for $n=0$, the integral\ \lsuy\ was  calculated previously 
in the work\ \dotse\ (see also Ref.\ \gr). 

\subsec{Renormalization group resummation}

Here  we  discuss the
short-distance expansion of the 
correlator \uytaa\ for $\beta^2$ sufficiently close to unity.
For this purpose, it is convenient to use the notation
\eqn\kaaaiu{\epsilon=1-\beta^2\ll 1\ .}
Our previous   CPT analysis
suggests the following expansion for
the structure function ${\Bbb C}_{\bf I}$:
\eqn\yut{{\Bbb C}_{\bf I}(r)=
r^{2d}\ \bigg\{\,
1+\sum_{k=1}^{\infty}c_k\, \big(\mu r^{2\epsilon}\big)^{2k}\, \bigg\}\ ,}
where
the coefficients $c_k$ are given by certain $4k$-fold 
Coulomb-type integrals.
Evidently, this   
expansion cannot be directly applied in the limit $\epsilon\to 0$,
where the perturbation $\cos(\beta\varphi)$ of the
Gaussian action\ \ooiy\ becomes marginal.
However, being expressed as a
function of the scaling distance\ $Mr$,
the structure
function ${\Bbb C}_{\bf I}(r)$ should admit the following form:
\eqn\uytra{{\Bbb C}_{\bf I}(r)=
\ Z_{n,\omega}\ {\Bbb C}^{(ren)}_{\bf I}(Mr)\ ,}
where  the  $r$-independent  renormalization  constant
$Z_{n,\omega}$ absorbs all divergences at  $\epsilon = 0$
and renders the renormalized  structure  function
 $ {\Bbb C}^{(ren)}_{\bf I}$ finite
in  this limit. The divergences of the renormalization constant
$Z_{n,\omega}$ should be directly related to the singularities of
${\Bbb C}^{(ren)}_{\bf I}$ at  $Mr=0$; they
point out that the power law  asymptotic behavior\ \cptres\ is
modified by logarithmic corrections  at  $\epsilon=0$.

In order to explore
the short-distance  behavior for  $\epsilon\ll 1$,
it is convenient
to return to the fermion description.
Being essentially the corresponding structure function in the
renormalizable QFT\ \suthirring,  ${\Bbb C}_{\bf I}(r)$  obeys the
Callan-Symanzik equation. Therefore  it  can be written in the form:
\eqn\kdidu{{\Bbb C}_{\bf I}(r)=
r^{-2d}\ \exp\bigg\{-2\, \int^{r}_0{dr\over r}\, \big(\, \Gamma_g-
d\, \big)
\, \bigg\}\ .}
Here the function $\Gamma_g$ is supposed to have a
regular power series expansion in
terms of the running coupling constants
$g_{\parallel,\perp}=g_{\parallel,\perp}(r)$:
\eqn\uytr{\Gamma_g=\sum_{l,k=0}^{\infty}
\gamma_{lk}\ g^l_{\parallel}\, g^{2k}_{\perp}\ .}
Notice that only even powers of the coupling $g_{\perp}$ appear
in this expansion\ (see footnote $\# 3$).
In writing\ \kdidu,  we use the normalization condition\ \norm, and
take into account that the UV limiting
value of $\Gamma_g$
coincides with the scale dimension\ \jsytr,
\eqn\ytre{\lim_{r\to 0}\Gamma_g=d\ .}
We  have  also assumed   that there is no resonance mixing of the operator
${\cal O}_{\omega\beta}^n$ with other fields,
 so it is  renormalized as a singlet.
One can easily  check that this is
indeed the case for the operators with  $|\omega|<{1\over 2}+{|n|\over 4}$.

Condition\ \ytre\ already encloses an important restriction
on the series\ \uytr. Indeed, using Eqs. \wer\ and \jsytr\
along with the condition that the line of UV stable fixed
points corresponds to  $g_{\perp}=0,$
one obtains
\eqn\krew{
\Gamma_g=\Gamma^{(0)}(g_{\parallel})+\Gamma^{(1)}
(g_{\parallel})\, g_{\perp}^2+
\Gamma^{(2)}(g_{\parallel})\, g_{\perp}^4+O( g_{\perp}^6)\ ,}
where
$$\Gamma^{(0)}(g_{\parallel})={2\omega^2\over 1+
{g_{\parallel}\over 2}}+
{n^2\over 8}\ \Big(1+{g_{\parallel}\over 2}\Big)\ .$$

The  values of the other coefficients $\gamma_{l,k\ge 1}$ appearing  
in\ \uytr\
essentially depend on the choice of a renormalization scheme, i.e
on the precise specification of
the running coupling constants.
The latter obey the RG equations
\eqn\RG{\eqalign{ r\,
{dg_{\parallel}\over dr}={g_{\perp}^2\over
f_{\parallel}(g_{\parallel}, g_{\perp})}\cr
r\, {dg_{\perp}\over dr}=
{g_{\parallel}g_{\perp}\over f_{\perp}(g_{\parallel}, g_{\perp})}\ .}}
Perturbatively, $f_{\parallel}(g_{\parallel}, g_{\perp})$
and $f_{\perp}(g_{\parallel}, g_{\perp})$
admit loop expansions as power series
in $g_{\parallel}$ and $g_{\perp}$.
In this work,
we will use the scheme introduced by Al.B. Zamolodchickov\ \refs{\Zms,
\Zmss}.
He showed that  under a suitable diffeomorphism
in  $g_{\parallel}$ and $g_{\perp}$, the functions  $f_{\parallel}$
and $f_{\perp}$ can be chosen to be equal to each other, and furthermore,
to be equal to
\eqn\yetr{f_{\parallel}=f_{\perp}= 1+{g_{\parallel}\over 2}\ .}
With this choice for the $\beta$-function,
the RG equations\ \RG\ can be integrated.
To do this, we note that
this system of differential equations
has a first integral,  the numerical value of which
is determined through the condition\ \wer,
\eqn\uyt{g_{\parallel}^2-g_{\perp}^2=(2\rho)^2\ .}
Using\ \uyt, \kaaaiu\ and\
\betag,  equations\
\RG\ are solved as
\eqn\gpargper{g_{\parallel}=
2\rho\ {1+q\over 1-q}\ ,\ \ \ \ g_{\perp}=\rho\ {4\sqrt{q}\over 1-q}\ ,}
where
\eqn\juyt{ q\ \Big({1-q\over \rho}\Big)^{-2\epsilon}=
({r\Lambda})^{4\epsilon}\, .}
The normalization
scale $\Lambda$ is
another integration constant
of the system \RG.
It is of the order of the physical mass scale and
supposed to have
a regular loop expansion,
\eqn\jus{\Lambda=M\ \exp\big(\tau_0+\tau_1\rho+ \tau_2\rho^2+
\ldots\big)\ .}
It should be noted that the even coefficients $\tau_0,\, 
\tau_2,\ldots$ are
essentially ambiguous and can be chosen at will.
A variation of these coefficients corresponds to a smooth
redefinition of the
coupling constants which does not affect the $\beta$-function.
By contrast, the odd constants $\tau_{2k+1}$ are unambiguous and
precisely specified once the
form of the RG equations is fixed. It is possible to show\ 
\refs{\Zms,\Zmss}\
that the odd constants vanish in  the Zamolodchikov's scheme:
$$\tau_{2k+1}=0\ \ \  \ \ \ \  (k=0,1\ldots)\ .$$

Once the coefficients $\tau_{2 k}$  in\ \jus\
are chosen,
the running   coupling constants are completely specified,
and all  coefficients in the power series expansion\ \uytr\ are
determined  unambiguously. They can be explicitly calculated by
comparing the CPT result\ \cptres\ with the  form\ \kdidu. From\ \kdidu,
$$\Gamma_g=-{1\over 2}\ r\, \partial_r \log\big(
{\Bbb C}_{\bf I}\big) $$
and, as it follows from the general 
CPT expansion\ \yut\ and the definition\ \juyt\ of $q$, the function
$\Gamma_g$
can be expanded  in 
powers of $q$. Explicitly, using the CPT result\ \cptres,
\eqn\kisuy{\Gamma_g=d-2\, \epsilon\ \Big( {\sqrt{\rho} \over
\Lambda } \Big)^{4\epsilon}\, \mu^2\ J_n\big(2\omega(1-\epsilon),
2\epsilon-2\big)\ q+O(q^2)\ . }
Moreover, the coefficients in this expansion
are power  series in $\rho$.
For example, using Eqs. \kytre\ and
\jus, it is easy to show that
\eqn\msnst{
{\pi\mu\over \epsilon}\, \Big( { \sqrt{\rho} \over
\Lambda} \Big)^{2\epsilon}= \exp\bigg\{-2{\bar \tau}_0 \rho+
\Big(
2{\bar \tau}_0
-{1\over 2}\Big)\rho^2-\Big(2\tau_2+2 {\bar \tau}_0
-{2\over 3}\, \zeta(3)-{1\over 2}
\Big)\, \rho^3+
O(\rho^4)\, \bigg\}\, .}
Here and after, we set for  convenience
\eqn\merr{ e^{\tau_0}=\sqrt{\pi\over 8}\ 
\ e^{\gamma_E+{\bar  \tau}_0}\ ,}
where $\gamma_E=0.5772\ldots $ is  the Euler constant.
The integral $J_n\big(2\omega(1-\epsilon),
2\epsilon-2\big)$ appearing in \kisuy\ can 
also be expanded in powers of $\rho$, using
$\epsilon = \rho/(1+\rho)$.
In  Appendix A, we quote the first 
few terms in the expansion of $J_n(a,c)$ \lsuy\
around $c=-2$, which are obtained through
the use  of\ \fsre. From this expansion, 
it is easy to obtain the expansion
of $J_n\big(2\omega(1-\epsilon),
2\epsilon-2\big)$ in powers $\rho$.
Then, one can  compare the CPT expansion 
of $\Gamma_g$ in $q$ and $\rho$ \kisuy\ with
the  corresponding expansion\ \uytr\ coming from the RG
analysis (where of course
one should expand $g_\parallel$ and $g_\perp^2$
in  $q$ and $\rho$ from \gpargper).
This determines  the  coefficients $\gamma_{l,1}$
for $l=0,1,2$.
If we want an expression valid to order $g^4$, we
need one more coefficient: $\gamma_{0,2}$.
In principle, it
can be obtained from the
expansion in $\rho$ of the coefficients $c_2$ 
in the series\ \yut.
In Section 5, we describe a way to find $\gamma_{0,2}$ without 
the cumbersome   calculation
beyond the lowest CPT  order.

In order to simplify the form of the structure function \kdidu,
it is convenient, instead of using the coefficients $\gamma_{l,k}$,
to parametrize the first few
terms of the power series expansions
$\Gamma^{(1,2)}(g_{\parallel})$
\ \krew\  as:
\eqn\vder{\eqalign{
&\Gamma^{(1)}(g_{\parallel})=
-{1\over 1+{g_{\parallel}\over 2}}\ \ \bigg\{\, {n^2\over 32}-{u_1
\over 2}+
v_1\, g_{\parallel}+\Big(v_2-{3u_2\over 2}\Big)\, g_{\parallel}^2
+O(g_{\parallel}^3)\, \bigg\}\ ,
\cr
&\Gamma^{(2)}(g_{\parallel})=
-{v_2\over 2}+
O(g_{\parallel})\ . }}
The explicit values of the coefficients
$u_1$, $u_2$, $v_1$ and $v_2$ in\ \vder\ are given in
Appendix B.

Let us
substitute\ \krew\ and \vder\
into  Eq. \kdidu.
The RG  flow equations\ \RG\ allow one to evaluate  the integral
and to write  the structure function  in the form\ \uytra\ with
\eqn\uytt{\eqalign{{\Bbb C}_{\bf I}^{(ren)}=&
 (Mr)^{-4\omega^2-{n^2(1+\rho^2)/ 4}}\ \
\big( g_{\perp}^2\big)^{\omega^2-{n^2(1-\rho^2)/ 16}}\times\cr & e^{-u_1
g_{\parallel}-u_2 g^3_{\parallel}}\ \Big(1+
g_{\perp}^2(v_1+v_2 g_{\parallel})+O(g^4)\, \Big)\ ,}}
and
\eqn\uytre{Z_{n,\omega}=M^{2d}\ \Big(2^{\rho+1}\sqrt{\rho}\,
e^{\tau_0\rho+\tau_2\rho^3+\ldots}
\Big)^{{n^2/ 2}-2d}\ \ \
 e^{2\rho u_1+(2\rho)^3 u_2+\ldots}\ .}
Notice that 
the transformation
\eqn\ndtr{
Z_{n,\omega}\to e^{w_0+w_1(2\rho)^2+w_2(2\rho)^4+\ldots}\ \ Z_{n,\omega}
\ ,}
where the  series contains only even powers of $\rho$ with
arbitrary coefficients $w_k$,
accompanied by
the transformation
$${\Bbb C}_{\bf I}^{(ren)}\to  e^{-w_0-w_1(g_{\parallel}^2-
g_{\perp}^2)-w_2(g_{\parallel}^2-
g_{\perp}^2)^2+\ldots}\ \ {\Bbb C}_{\bf I}^{(ren)} $$
does not affect the structure function  ${\Bbb C}_{\bf I}$\ \uytra\
due to relation\ \uyt.

Our prime interest 
in  this work is  the correlation function\ \terju.
For  $n=1$\ and\ $\omega={1\over 4}$, the  relations  obtained above 
lead to the following perturbative
expansion for the two-point fermion correlator in
the anisotropic  $SU(2)$-Thirring model:
\eqn\ksure{
\eqalign{\langle\, \Psi_{\sigma'}(x)\, 
&{\bar \Psi}_{\sigma}(0)\, \rangle=
{ Z_{\Psi}\delta_{\sigma'\sigma}\over 2\pi}\ \
{ \gamma_{\mu}x^{\mu}\over |x|^{2+{\rho^2\over 4}}}\ \
\big( g_{\perp}^2\big)^{\rho^2\over 16}\
\exp\bigg\{-{3\over 16}\ g_{\parallel}-{{\bar \tau}_{0}\over 32}\
g_{\parallel}^3\, \bigg\}\times\cr & \exp\bigg\{{3\over 16}\,
\Big({\bar \tau}_{0}-{1\over 4}\Big)\ g_{\perp}^2-{3\over 16}
\Big({\bar \tau}_{0}^2-
{1\over 6}\, {\bar \tau}_{0}-{1\over 16}\Big)\
g_{\parallel}
g_{\perp}^2
+O(g^4)\, \bigg\}\ ,}}
where
$$Z_{\Psi}=(4\rho)^{-{\rho^2\over 8(1+\rho)}}\ \bigg(M\sqrt{\pi\over 2}
\, \bigg)^{-{\rho^3\over 4(1+\rho)}}\ \exp\bigg\{
{3\rho\over 8}-{\gamma_{E}\over 4}
\, \rho^3+O(\rho^4)\bigg\}\ .$$
In Eq. \ksure, we  use the notation ${\bar \tau}_0$ defined
by \merr.

We now set $\rho=0$ and $g_{\parallel}=
g_{\perp}=g$ in\ \ksure\ to
obtain the perturbative expansion of the
scaling function $F$\ \piu\ for
the $SU(2)$-Thirring
model,
\eqn\ksureaa{
F^{(pert)}=
\exp\bigg\{-{3\over 16}\, g+{3\over 16}\, \Big({\bar \tau}_{0}-
{1\over 4}
\Big)\, g^2-{3\over 16}\, \Big({\bar \tau}^2_{0}-{1\over 16}
\Big)\, g^3+O(g^4)\, \bigg\}\ .}
Here the running coupling 
constant $g$ solves the equation
\eqn\ncbt{-g^{-1}+{1\over 2}\, \ln(g)=\ln\Big(\, \sqrt{\pi\over 2}\,
e^{\gamma_E+{\bar \tau}_{0}}\, Mr\,
\Big)\ ,}
which is  the limit $\rho=0$ of Eqs. \gpargper\ and \juyt.

Let us stress here that, if the perturbation series could be summed, then
the function $F$ should not
depend on the  auxiliary parameter ${\bar \tau}_{0}$:
$${\partial F\over \partial {\bar \tau}_{0}}=0\ .$$
This is, however, not true if we truncate the series\ 
\ksureaa\ at  some order $N$
(for instance, if one  leaves
only the terms explicitly written in \ksureaa). In this case,
$${\partial \over \partial {\bar \tau}_{0}}\, 
F^{(pert)}_N=O\big(g^{N+1}\big)\ ,$$
where  the truncated series is  denoted  by\ $F^{(pert)}_N$.
In fitting numerical data with\ \ksureaa, we may treat
${\bar \tau}_0$ as an optimization parameter, allowing us 
to minimize or at least develop a feeling for
the effects of the remainder of the series. Similar ideas
have been discussed for QCD in Ref. \QCD.  

It may be worth  mentioning that
Eq.\uytt, along with  explicit values of the coefficients
quoted in  Appendix B, allows one to immediately determine
the short-distance expansion of some other  conventional correlators in
the (anisotropic)  $SU(2)$-Thirring model.
For example, since the sine-Gordon field $\varphi$\ \ksiy\
itself can be defined by the relation
$$\varphi=-i\, 
{\partial\over \partial a} {\cal O}_{a}^{n}\Big|_{n=0\atop a=0}\
,$$
and the spin current $J^3_{\mu}$\ \currs\ is bosonized as
\eqn\mnfsrde{J^3_{\mu}={\beta\over 2\pi}\, \partial_{\mu}\varphi\ ,}
we can use\ \uytt\ to obtain the short-distance expansion
of the current-current correlator.
For the $SU(2)$-Thirring model\ \thirringi\ one has,
\eqn\msdtew{\langle\, J^A_{\mu}(x)\, J^B_{\nu}(0)\, \rangle=
{\delta^{AB}\over \pi^2\, |x|^2}\
\bigg\{\, \Big(\, \delta_{\mu\nu}-
 {2x_{\mu}x_{\nu}\over |x|^2}\Big)\
I_1+{\delta_{\mu\nu}}
\ I_2\bigg\}\ ,}
where
$$\eqalign{&I_1=1-{g\over 2}+ \Big({\bar \tau}_0+{1\over 4}
\Big)\, {g^2\over 2}-{\bar \tau}_0({\bar \tau}_0+1)\,
{g^3\over 2}
+\Big({{\bar \tau}_0^3\over 2}+
{\bar \tau}_0^2+{{\bar \tau}_0\over 4}-
{13\over 128}-{7\over 16}\, \zeta(3)
\Big)\, g^4+O(g^5) ,\cr
&I_2={g^2\over 4}\, 
\bigg\{1-2\, {\bar \tau}_0\, 
g+{\bar \tau}_0(3\, {\bar \tau}_0+1)\, g^2-
\Big(4\, {\bar \tau}_0^3+{7\over 2}\, {\bar \tau}_0^2
+{{\bar \tau}_0\over 2}-{13\over 16}-{7\over 2}\, \zeta(3)
\Big)\, g^{3}+O(g^4)\bigg\}
\, ,}$$
and the running coupling constant $g$ is the same as in\ \ncbt.

\newsec{Perturbative expansion of the 
momentum-space correlation function}

\subsec{ Large-momentum asymptotics} 

Perturbative calculations of 
fermion Green's functions in renormalizable
2D models with four-fermion interaction    are widely covered in the
literature.
The  results in this domain are  usually
expressed in momentum space.  Hence it 
seems appropriate at this point
to adapt the calculation of the previous section to the momentum-space
fermion correlator, giving a large-momentum expansion.

The RG analysis performed in the previous
section can be applied in essentially the same way to the Fourier
transform of the fermion correlator \terju:
\eqn\Fmomdef{
\int d^2x\, e^{-ip x}\ 
\big\langle\, \Psi_{\sigma}(x)\, 
{\bar \Psi}_{\sigma'}(0)\, 
\big\rangle = -i\, \delta_{\sigma\sigma'}\ 
{\gamma^\mu p_\mu \over p^2}\  \ {\tilde F}(p^2)\ .
}
Here and after we use the notation $p^2=p^{\mu}p_{\mu}$.
{}From the result of CPT, \cptres, one
immediately obtains the large momentum expansion of 
this Fourier transform:
\eqn\Fmom{
{\tilde F} = Q(d_\Psi)\, (p^2)^{d_\Psi-{1\over 2}}\, \bigg\{ 1 +
{Q(d_\Psi-2\epsilon)\over Q(d_\Psi)}\, J_1(\beta^2/2,-2\beta^2)\ 
 \mu^2(p^2)^{-2\epsilon}+
O\big((p^2)^{-4\epsilon},p^{-2}\big) \bigg\},
}
where
$$
Q(a) =2^{1-2a}\ {\Gamma({3\over 2}-a)\over \Gamma({1\over 2}
+a)}\ ,$$
and 
$$d_\Psi = {1\over 2} +{\rho^2\over 4(1+\rho)} $$
is the scale dimension of the fermion field.
The factor 
$Q(d_\Psi-2\epsilon)/Q(d_\Psi)$ is 
essentially the only source of differences
between the RG treatments in coordinate 
space and in momentum space. The RG analysis 
in momentum space goes as in the previous section. 
The perturbative
part in $\mu$ of  $ {\tilde F}$
obeys the Callan-Symanzik equation, so it can be written as
\eqn\Fmomcs{
{\tilde F}^{(pert)}=
Q(d_\Psi)\ (p^2)^{d_\Psi-{1\over 2}}\  
\exp \bigg\{ -\int_{p^2}^\infty {ds\over s}\,
\big({\tilde\Gamma}_g - d_\Psi\big) \bigg\}\ ,
}
where the function ${\tilde \Gamma}_g$   admits a power
series expansion in terms of the momentum-space 
running coupling constants
$g_{\parallel,\perp}=
g_{\parallel,\perp}(p^2)$\ depending on the Lorentz invariant $p^2$: 
\eqn\gam{
{\tilde \Gamma}_g = 
\sum_{l,k=0}^\infty {\tilde\gamma}_{l,k}\ 
g_{\parallel}^l g_{\perp}^{2k}\ .}
Notice that, with some abuse of
notations, we  use here the same symbols $g_{\parallel,\perp}$
for the momentum-space running couplings as
we used for the coordinate-space  running couplings.
In order to fix the  coefficients in\ \gam, 
we have to choose a renormalization scheme.
Substituting $r$ by $1/\sqrt{p^2}$ in\ \juyt\  defines 
Zamolodchikov's scheme
in momentum space. 
It is a simple matter to repeat 
the steps of the previous section in
order to determine the first few coefficients ${\tilde\gamma}_{l,1}$ 
in \gam.
Just compare the logarithmic derivatives of the
expressions \Fmom\ and \Fmomcs;
the only difference is that the factor $Q(d_\Psi-2\epsilon)/Q(d_\Psi)$
in \Fmom\ will have to be expanded in $\rho$, giving non-trivial
contributions. As for the coefficients ${\tilde\gamma}_{l,2}$, one
would in principle need the next order in CPT. However, again as
in the previous section, 
it is possible to determine ${\tilde\gamma}_{0,2}$
without this calculation, as described in the next section.
{}From these coefficients,
and from the form of the RG flow equation, 
one can evaluate the integral in \Fmomcs\ 
and obtain the asymptotic  behavior 
of the
two-point function in the Euclidean region at
$p^2\to+\infty$.
We  quote here 
the  result in the case of the $SU(2)$-Thirring model, 
\eqn\Fmomsuth{
{\tilde F}^{(pert)}=
\exp\bigg\{-{3\over 16}\, g+
{3\over 16}\, \Big({\tilde \tau}_{0}-{1\over 4}
\Big)\, g^2-{3\over 16}\, \Big({\tilde \tau}^2_{0}
-{1\over 16}
\Big)\, g^3+O(g^4)\, \bigg\}\ .
}
Here
\eqn\scalemom{
-g^{-1}+{1\over 2}\, \ln(g)=\ln\big(\, 
\sqrt{2\pi} M\, e^{{\tilde \tau}_{0}}/
\sqrt{p^2}\,  
\big)\ ,
}
and ${\tilde \tau}_0$ is 
an arbitrary parameter which can be chosen at will.
Notice the strong similarity between \Fmomsuth\ and \ksureaa.

We also quote here
the corresponding  function\ ${\tilde\Gamma}_g$\ \gam\ in the case
$g_{\parallel} = g_{\perp}$:
\eqn\anmomsuth{
{\tilde\Gamma}_g = {1\over2} + {3\over32}\, 
g^2 -{3\over16}\, {\tilde \tau}_0\ g^3 + {3\over32}\, \Big(
\, 3\, {\tilde \tau}^2_0 + {\tilde \tau}_0
 - {3\over16} \Big)\ g^4 + O(g^5)\ .
}

\subsec{Comparison with the four-loop conventional
perturbation calculations}

In\ \AliGracey, the anomalous dimension for
the fermion field in the 
${\overline {\rm MS}}$ 
scheme was found to fourth order for
a general non-abelian Thirring model (see also\ \BennettGracey\
and references therein for a discussion of various
aspects of dimensional regularization 
in the non-abelian Thirring model and for results to lower order).
In contrast, we have calculated, in coordinate space, the two-point
functions
of more  general operators, 
including the fermion fields, in the particular
case of the $SU(2)$-Thirring model 
(and an anisotropic deformation of it),
and we have sketched the equivalent calculation in momentum space for
the fermion fields. We would now like to compare Eq.\Fmomsuth\
with the $SU(2)$ case of the Ali-Gracey result
\AliGracey.
In order to perform the comparison,
we need to find the relation between
our running  coupling constant $g$ and theirs, 
which will be  denoted $g_{AG}=-\lambda\, $ \foot{
Notice that in\ \AliGracey, the coupling constant $g_{AG}$
is assumed to be 
negative, so $\lambda>0$, 
which agrees with the sign of our coupling constant $g$.},
and then
find the
relation between our function ${\tilde\Gamma}_g$\ \Fmomcs\
and their anomalous dimension, which
we will denote $\gamma_\lambda$.

The coupling $\lambda$
corresponds to the ${\overline {\rm MS}}$ scheme; 
the associated $\beta$-function
was found in\ \BennettGracey\ to 
fourth order:
\eqn\hdytr{2\, p^2\,
 {d\lambda\over d p^2}=
\beta_\lambda = -{\lambda^2\over\pi} + {\lambda^3\over 2\pi^2} - 
{83\over 128\pi^3}\ \lambda^4 + O(\lambda^5)\ .}
By comparison, in the scheme  that we use, the $\beta$-function\ 
\RG,\ \yetr\ is
\eqn\jduyt{
2\, p^2\,
 {dg\over d p^2}=
\beta_g =-{g^2\over 1+g/2}= -g^2+{g^3\over2} - {g^4\over4} + O(g^5)\ .}
The difference in the factor multiplying the square of the coupling
in these two expressions 
results only from a different normalization of the coupling 
in the action (see Eq.\thirringi).
The relation between the couplings $g$ and $\lambda$ that 
corresponds to these different $\beta$-functions is
\eqn\ggt{{\lambda\over\pi} = 
g - \tau \,g^2 + 
\Big( \tau^2 + {\tau\over2} + {51\over128} \Big)\,  g^3 + O(g^4)\ .}
Here $\tau$ is some numerical factor which cannot
be determined by comparing the $\beta$-functions: its variation
modifies  the choice of the normalization scale and
doesn't affect the $\beta$-beta functions.
The normalization scale for the ${\overline {\rm MS}}$ scheme
is defined by imposing the following condition on 
the subleading asymptotics of the solution of the
RG  flow equation\ \hdytr:
\eqn\sca{
{\lambda\over\pi}={1\over \ln\big(\sqrt{p^2}/
\Lambda_{{\overline {\rm MS}}}
\big)}+
{1\over 2} \ { \ln\ln(\sqrt{p^2}/\Lambda_{{\overline {\rm MS}}})\over
\ln^2(\sqrt{p^2}/\Lambda_{{\overline {\rm MS}}})}+O\bigg(
{ \ln^2\ln(\sqrt{p^2}/\Lambda_{{\overline {\rm MS}}})\over
\ln^3(\sqrt{p^2}/\Lambda_{{\overline {\rm MS}}})}\bigg)\ .}
(This implies that the term 
$O\big(1/\ln^2(\sqrt{p^2}/\Lambda_{{\overline {\rm MS}}})\big)
$ does not appear in the
expansion of $\lambda$.)
From\ \scalemom, \ggt\ and \sca, we find that
\eqn\kduyt{
\Lambda_{\overline {\rm MS}} = \sqrt{2\pi}M\ e^{{\tilde \tau}_0-\tau}\ .}

In\  \AliGracey, the perturbative
part of the  function\ ${\tilde F}$ \Fmomdef\ was calculated
up to the  overall multiplicative  normalization 
to third order in $\lambda$.
The result can be  written in
the following form
$${\tilde F}^{(pert)} \propto {1\over h_{\lambda}}\ 
\exp\bigg\{-{1\over 2}
\int^{p^2}  {ds\over s}\
{\gamma}_\lambda \bigg\} \ ,$$
where the function $h_{\lambda}$ and the anomalous dimension 
$\gamma_\lambda$
were given in\ \AliGracey\ to   fourth order in 
$\lambda$
for the Thirring model with a general non-abelian  symmetry.
In the particular case of the $SU(2)$-symmetry,
they specialize to
\eqn\bdgr{
h_{\lambda} = 
1+{15\over 128\pi^2}\ \lambda^2 - {11\over 512\pi^3}\, \lambda^3  +
{3(80\zeta(3)-511)\over 32768\, \pi^4}\ \lambda^4+ O(\lambda^5)\, ,}
and
\eqn\msht{{\gamma}_\lambda=
-{3\over 16\pi^2}\  \lambda^2 + {15\over 64\pi^3}\ \lambda^3+
{3\over 1024\pi^4}\  \lambda^4 + O(\lambda^5)\ .}
Comparing\ \Fmomcs\ in the case
$\rho = 0$ \  with 
the above expressions, one has
the following relation:
\eqn\jsuytnl{{\tilde\Gamma}_g =
{1\over2} - {{\gamma}_\lambda\over2} -
{\beta_\lambda \over2}\, {d \over d{\lambda}}\, 
\log({h}_{\lambda})\ .}
Using Eqs. \ggt-\msht, one can check that
our result\ \anmomsuth\  agrees with\ \jsuytnl, provided that
\eqn\ggtt{\tau ={\tilde \tau}_0\ .}
Notice that the relation between  the normalization scale
$\Lambda_{\overline {\rm MS}}$ and
$M$,
\eqn\sdnaytlm{\Lambda_{\overline {\rm MS}} = \sqrt{2\pi}M\ ,}
which is a consequence of\ \kduyt\ and\ \ggtt, was previously 
found in Ref. \Woynarovich.

\newsec{Long-distance  behavior}

Here we concentrate on the 
long-distance behavior 
of   Schwinger's function\ \uytaa\ for $n=1$ 
and $\omega=1/4$.
Let us recall that for ${1\over2 }<\beta^2\leq 1$, there are only 
solitons and antisolitons in the
spectrum of the sine-Gordon model. We will denote them
by $A_-$ and $A_+$ respectively. 
The conservation of the topological charge,
$${\beta\over 2\pi}\ \int_{-\infty}^{\infty}d{\rm x}\, 
\partial_{\rm x}\varphi\ ,$$
implies that the non-vanishing form factors of the operator ${\cal
O}_{\beta/4}^{+1}$ are of the form
\eqn\higher{
\langle\, vac \mid {\cal O}_{\beta/4}^{+1}(0)\mid A_{-}(\theta_{1})\cdots
A_{-}(\theta_{N+1})\,  
A_{+}(\theta_1')\cdots A_{+}(\theta_N')\, \rangle\ ,}
where $\theta_i$ and $\theta_j'$ denote rapidities of solitons and
antisolitons respectively.  Up to an overall normalization, all these 
form factors can be written down in closed form, as certain 
$N$-fold integrals \refs{\sm,\lik,\drew}. The spectral decomposition
for the correlation function\ \terju\ then gives
\eqn\dfre{\eqalign{&F_{{1/4}}^{(1)}(r)=
\int_{-\infty}^{+\infty}{d\theta\over 2\pi}\ e^{-Mr \cosh(\theta)}\,
|\langle\, \, vac \mid {\cal O}_{\beta/4
}^{+1}(0)\mid A_{-}(\theta)\,\rangle|^2+
{1\over 3!}\int_{-\infty}^{\infty}
{d\theta_1d\theta_2d\theta_3\over (2\pi)^3}\times\cr &
 e^{-Mr\sum_{k=1}^3
\cosh(\theta_k)} 
\sum_{
\sigma_1+\sigma_2+\sigma_3=-1}
|\langle\, vac\,\mid\,
{\cal O}_{\beta/4}^{+1}(0)
\mid A_{\sigma_1}(\theta_1)A_{\sigma_2}(\theta_2)
A_{\sigma_3}(\theta_3)
\,\rangle |^2+\ldots\ ,}}
where the dots
stand for the five-particle and higher contributions,  which are of
the order of $e^{-5Mr}$.
The  long-distance asymptotic behavior of the correlation
function is dominated by the contribution of the one-particle
states,
$$\langle\, \, vac \mid {\cal O}_{\beta/4
}^{+1}(0)\mid A_{-}(\theta)\,\rangle=\sqrt{{\bf Z}_{1}(\beta/4)}\ 
e^{{1\over 4}\, (\theta+{i\pi\over 2})}\ ,$$
and has an especially simple form,
\eqn\liy{F_{{1/ 4}}^{(1)}(r) =
{\bf Z}_1(\beta/4)
\ \bigg\{\, {e^{-Mr}\over \sqrt{2\pi Mr }}+O(e^{-3Mr})\, \bigg\}\ .}
Here we use the notation\ ${\bf Z}_n(a)\ (a=\omega\beta)$ from 
work \LZtopff\ for  the field-strength renormalization 
which controls the long-distance
asymptotics of  the  correlation  function\ \uytaa.
Let us stress here that the overall multiplicative
normalization of the field\ ${\cal O}_{\beta/4}^1$ was already
fixed by the condition\ \norm, hence
the constant\ ${\bf Z}_1 (\beta/4)$ is totally unambiguous.
In \LZtopff,
the following explicit  formula for ${\bf Z}_n (\omega\beta)$
was  proposed:
\eqn\lsidiiu{\eqalign{&{\bf Z}_n(\omega\beta)=   \Big({{\cal C}_2
\over 2\,  {\cal C}_1^2}\Big)^{n\over 2}\ \ \Big(
{ {\cal C}_2\over 16\rho}\, \Big)^{-{n^2\over 4}}\ \
\bigg[\, {\sqrt{\pi} M \Gamma({3\over 2}+
{1\over 2\rho})\over 2\, \Gamma(1+{1\over 2\rho})}\, \bigg]^{2d}\times
\cr &
\exp\bigg[\, \int_{0}^{\infty} {dt\over t}\
\Big\{\, {\cosh(4\omega t)\, e^{-(1+\rho) nt}-1\over 2\, \sinh( t)
\sinh\big(\, (1+\rho) t\, \big)\, \cosh(t\rho)}+{n\over 2\, \sinh(t)}
-2d\ e^{-2 t}
\, \Big\}\, \bigg]\ .}}
In this formula,  $d$ is the scale dimension\ \jsytr\ and
the constants\ ${\cal C}_1, \ {\cal C}_2$ read
$$\eqalign{&{\cal C}_1={ 2^{-{5\over 12}}\, 
e^{1\over 4}\, \Gamma({1\over 4})\over
\sqrt{\pi}\, A_G^{3}}\ 
\exp\bigg\{\,
\int_{0}^{\infty} {dt\over t}\ {\sinh^2({t\rho\over 2})\,
e^{-t}\over 2\, \cosh^2(t\rho)\, \sinh(t)}\,\bigg\}\, ,\cr
&{\cal C}_2= {\Gamma^4({1\over 4})\over
4\pi^3}\
\exp\bigg\{\, -2\, \int_{0}^{\infty} 
{dt\over t}\ {\sinh^2({t\rho\over 2})\,
e^{-t}\over \cosh(t\rho)\, \sinh(t)}\,\bigg\}\ ,}$$
where  $A_G=1.282427\ldots$ is the Glaisher constant.

We do  not write  down explicitly the general formula
for the three-particle contribution
in\ \liy\ 
because it is a rather mechanical substitution of
relations presented in \LZtopff. 
(For $\beta^2=1$ the corresponding formulas can be found 
in Appendix C.)
Here we make the following observation concerning the
$\beta^2\to 1$ limit. 
The examination of\ \dfre\ based on
explicit formulas for the form factors shows
that the function $
\big[{\bf Z}_1 (\beta/4)\big]^{-1}\, F_{{1/ 4}}^{(1)}$
admits an asymptotic power series expansion in terms of the variable
$\rho^2$.
In other words, all  divergences at  $\rho^2\to 0$
of $F_{{1/ 4}}^{(1)}$, 
considered as a
function of the variables  $\rho^2$ and $Mr$, are  absorbed by the
normalization constant ${\bf Z}_1(\beta/4)$.
Using Eq. \lsidiiu, one can check
that the constant ${\bf Z}_1(\beta/4)$ admits exactly the same
type of singular  behavior at $\rho^2=0$
as the constant $Z_{n,\omega}$\ \uytre\ 
for $n=1,\, \omega=1/4$, and also that
\eqn\wqas{{{\bf Z}_1(\beta/4)
\over  Z_{1,1/4}}=2^{-{1\over 3}}\ \sqrt{\pi}\,  e^{-{1\over 4}}\ A_G^3\ 
\exp\big(w_1\, \rho^2+O(\rho^3)\big)\ .}
The explicit form of the coefficient $w_1$ is not
essential here. What is  important is that
the linear term in $\rho$ does not appear in the expansion\ \wqas.
This observation  can be immediately generalized and
checked for  any $n$ and $\omega$. Furthermore,
we expect that
\eqn\lskit{\log\Big({{\bf Z}_n(\omega\beta)
\over  Z_{n,\omega}}\Big)=\sum_{k=0}^{\infty}
w_k\ \rho^{2k}+O(\rho^{\infty})\ ,}
where ${\bf Z}_n(\omega\beta) $ 
is  the normalization  constant\ \lsidiiu.
In other words,  by means of the transformation\ \ndtr\ with  
properly chosen coefficients $w_k$,
the constant $Z_{n,\omega}$ in\ \uytra\ can be set to be equal
(in a sense of formal power series)
to ${\bf Z}_n(\omega\beta)$.
At the moment, we do not have a rigorous proof of\ \lskit.
But it leads to some interesting prediction to  be checked.
As was already mentioned,  the calculations performed
in the leading CPT order determine only
the combination $v_2-3u_2/2$, but do not fix the individual values of 
the  coefficients
$u_2$ and $v_2$ in the series\ \uytt. 
Accepting\ \lskit,  one can immediately find  the values 
of the coefficients $u_2$ (see Appendix B).
In the case   $n=1,\, \omega={1\over 4}$, it 
allows one to extend the perturbative
expansion\ \ksureaa, as well as  the equivalent  expansion
\Fmomsuth, to order  $g^3$.
As was discussed in Section 4,  Eq. \Fmomsuth\ 
is in a complete 
agreement with the result of  four-loop perturbative 
calculations from\ \AliGracey.
This in fact shows that the $\rho^3$-term  really is absent in
the series \wqas.

\newsec{Spectral density}

The spectral density is an 
important quantity related to the two-point function and its
analytical structure in momentum space. 
It is often what is measured in actual condensed matter
experiments\ \refs{\exper,\Fab}, and it 
allows one to completely reconstruct 
the two-point function. 
In this section,
we discuss the properties of the
spectral density in the $SU(2)$-Thirring model.

The spectral decomposition of
the  fermion Green's  function  yields the following form for
the function\ ${\tilde F}$\ \Fmomdef: 
\eqn\jsutyr{{\tilde F}(p^2)= 
1-\int_{M^2}^{+\infty} ds\  {\Delta{\tilde F}(s) \over p^2+s}\ .}
The notation $\Delta{\tilde F}$ for the spectral density
reminds us that ${\tilde F}$, 
considered as a function of one complex variable $p^2$,
has
a branch cut in the Minkowski region $p^2<0$ starting
at $p^2=-M^2$,  and that
the spectral density can be recovered from
the discontinuity along  this cut:
\eqn\DeltaGdef{
\Delta{\tilde F}(s) = 
{1\over 2\pi i  }\  \big(\, {\tilde F}( e^{i\pi} s)-
{\tilde F}( e^{-i\pi} s )\, \big)\ .}
The easiest way to obtain the large $s$ asymptotics of
the spectral density  
is to use the expansion\ \Fmomsuth\ along with
knowledge of the analytical properties
of  the coupling constant $g$\ \scalemom\ as a function 
of the complex variable $p^2$.
Notice that $g$ can be expressed
in terms of  the  principal branch of the product log   (or Lambert)
function,  which gives the solution for $W$ in  
$W\, e^{W} = z$\ (see e.g. \CGHJK):
\eqn\couplingW{g=2\  W^{-1}\Big({p^2 e^{-2{\tilde \tau}
_0}\over \pi M^2}\Big)\ .}
The principal branch of the  $W$-function 
analytically maps the complex $z$-plane minus
the branch cut $z\in]-\infty,-e^{-1}]$
to the part of the complex 
$W$-plane enclosing the real axis and delimited by the curve
$\Re e\, W =
-\Im m\, W\,\cot(\Im m\, W)$ for $-\pi<\Im m\, W < \pi$.
The analyticity implies that the  power series
$$
\sum_{n=0}^{\infty}\, {1\over n!}\
\Big(i\phi\,  z\, {d\over dz}\Big)^n\  W(z)\Big|_{z=s} 
$$
converges for real positive $s>e^{-1}$ and $|\phi|\leq \pi$ and
coincides with $W(e^{i\phi} s)$.
Similar considerations are, of course, valid 
for the coupling constant $g$\ \couplingW. In particular,
for sufficiently large $s$,
$$g(e^{\pm i\pi} s)=\sum_{n=0}^{\infty}\, {1\over n!}\
\Big(\pm i\pi\,  p^2\, {d\over dp^2}\Big)^n\  g(p^2)\Big|_{p^2=s}\ .$$
This then
gives us, with
\Fmomsuth\ and the RG flow equation\ \jduyt,
the asymptotic expansion of the spectral density   for large $s$.
It can be written in the following form:
\eqn\slkkii{\Delta{\tilde F}(s)=-{g^2\over 2}\ \bigg\{1-{g\over 2}-
{\pi^2-1\over 4}\ g^2+O(g^3)\, \bigg\}\ {
\partial {\tilde F}^{(pert)}\over \partial g}\, \bigg|_{p^2=s}\ .}
Here the function ${\tilde F}^{(pert)}$ is given by \Fmomsuth\ and
$g$ is defined  by the equation\ \scalemom.

Now let us consider the threshold behavior of the spectral density.
According to the analyses of the previous section, the long-distance
asymptotic behavior  of the
scaling function $F$\ \piu\ is described by the expansion
\eqn\bsgr{F=F^{(1)}+F^{(3)}+O(e^{-5Mr})\ ,}
where
$$ F^{(1)}=C\ e^{-Mr}\ ,$$
with the constant
$$C=2^{-{5\over 6}}\ e^{-{1\over 4}}\  A_G^3=
0.921862\ldots\ .$$
The function $F^{(3)}$ in\ \bsgr\ gives  the three-particle
contribution to the correlation function. Using the definitions\ 
\piu, \Fmomdef\ and the above relation, one can obtain:
\eqn\zsfdr{
{\tilde F}(p^2)=C\ \bigg\{\, 1-{1\over \sqrt{1+p^2/M^2}}\, \bigg\}+
\ldots\ .}
Here  the dots stand for 
contributions of the massive multiparticle intermediate  states.
The last relation implies that the spectral 
density\ \DeltaGdef\ can be written as
\eqn\ksuaay{
\Delta{\tilde F}(s)={C\over\pi}\ {\Theta(s-M^2)\over \sqrt{s/M^2-1}}+
\Theta(s-9M^2)\ \Delta{\tilde F}^{(3)}(s)\ ,}
where
$$\Theta(s)=\cases{1\ \ {\rm for}\  s\geq 0\cr 0 
 \ \ {\rm for}\    s<0 }\ ,$$
and  $\Delta{\tilde F}^{(3)}$ is some function which contributes
to the spectral density only above the threshold $s=9M^2$.

\newsec{Numerics}

In Table 1 we present  results of numerical evaluation of the 
function $F$\ \piu\ as a function of the scaling distance
$Mr\ (r=|x|)$.
To estimate the short-distance behavior, we use the perturbative
expansion\ \ksureaa. As was already mentioned, 
the parameter ${\bar \tau}_0$
allows one to have control over the accuracy of
the truncated series, so we calculate\ \ksureaa\  
for two different values of\ $ {\bar \tau}_0:-0.25$ and
$+0.25$.
To determine the long-distance behavior of the function $F$, we use
the  formula\ \bsgr, where
the three-particle contribution
$F^{(3)}$ was obtained by means of Eq. \dfre\ along with
formulas for  the three-particle 
form factors quoted in \LZtopff\ 
(see Appendix C).
It is interesting to see
that the sum of the one- and three-particle contributions to $F$ 
is very near to unity at $r=0$ (to within 1\%), 
which indicates that this three-particle computation of the correlation
function is in fact accurate to about 1\% for  all distance scales 
(more accurate, of course,  for larger $r$).
Also, note that the crossover between the  long- and
short-distance asymptotics appears to be at the scaling distances
$Mr\sim 0.001-0.01$, where both asymptotics
coincide to within about $0.1\%$.
\midinsert
\centerline{
\noindent\vbox{\offinterlineskip
\def\tablerule{\noalign{\hrule}}
\halign{
\strut#&\vrule#\tabskip=1em plus2em&
   #&\vrule#&
   #&\vrule#&
   #&\vrule#& 
   #&\vrule#&
   #&\vrule#& 
   #&\vrule#
\tabskip=0pt
\cr\tablerule
&& $Mr$  &&\  $F^{(1)}$   && $F^{(3)}$ && $F^{(1)}+F^{(3)}$
&& $F^{(pert)}$  $({\bar \tau}_0=-0.25)$  && $F^{(pert)}$ 
$({\bar \tau}_0=0.25)$ &
\cr\tablerule
&& 0 && .921862 && .068 && .990 && 1.00000 && 1.00000 & \cr \tablerule 
&& .00001 && .921853 && .0553 && .9771 && .980129 && .980130 &\cr \tablerule 
&& .00005 && .921816 && .0522 && .9740 && .976311 && .976314 &\cr \tablerule 
&& .0001 && .921770 && .0504 && .9722 && .974192 && .974196 & \cr \tablerule 
&& .0002 && .921678 && .0483 && .9700 && .971674 && .971678 & \cr \tablerule 
&& .001 && .920941 && .0415 && .9624 && .963508 && .963520 & \cr \tablerule 
&& .002 && .920020 && .0375 && .9575 && .958435 && .958454 & \cr \tablerule 
&& .01 && .912689 && .0252 && .9379 && .939386 && .939460 & \cr\tablerule 
&& .025 && .899101 && .0168 && .9159 && .919294 && .919494 & \cr\tablerule 
&& .05 && .876902 && .0106 && .8875 && .894050 && .894547 & \cr\tablerule 
&& .075 && .855251 && .00738 && .86263 && .871796 && .872717 &\cr\tablerule 
&& .1 && .834135 && .00541 && .83955 && .850520 && .852013 & \cr\tablerule 
&& .15 && .793454 && .00317 && .79662 && .808380 && .811548 & \cr\tablerule 
&& .2 && .754757 && .00200 && .75676 && .765139 && .770842 & \cr\tablerule 
&& .25 && .717947 && .00131 && .71926 && .719980 && .729252 &\cr\tablerule 
&& .3 && .682932 && .000889 && .683822 && .672640 && .686654 &\cr\tablerule 
&& .35 && .649625 && .000617 && .650243 && .623153 && .643171 &\cr\tablerule 
&& .4 && .617942 && .000436 && .618379 && .571774 && .599063 &\cr\tablerule 
&& .45 && .587805 && .000313 && .588118 && .518942 && .554677 &\cr\tablerule 
&& .5 && .559137 && .000227 && .559365 && .465257 && .510405 &\cr\tablerule 
\noalign{\smallskip} }
}
}
\capt{Table 1. The scaling function $F$\ \oip, \piu.
The first  columns give 
the results of the long-distance expansion which
includes contributions of the 
one-, three- and one$+$three-particle states. The
data in the last two columns correspond to the 
perturbative expansion\  \ksureaa\ for\ 
the two different values of the auxiliary parameter ${\bar \tau}_0$. }
\endinsert

\bigskip

\centerline{\bf Acknowledgments}

\bigskip

S.L. is grateful to F. Essler and  A. Tsvelik for 
their helpful 
collaboration  during  the early stages of this work.
We  have also extremely   benefited  from
discussions with   A. Zamolodchikov and Al. Zamolodchikov.

This research is supported in part by DOE grant $\#$DE-FG02-96ER10919.
B.D. is supported in part by a NSERC Postgraduate Scholarship.

\vfill
\eject

\appendix{A}{}

In this appendix, we 
give the first few terms 
in the expansion of $J_n(a,c)$ \lsuy\ around $c=-2$.
The coefficients in this expansion 
involve standard functions of $a$, which could then easily be used
to obtain an expansion of
$J_n\big({2\omega\over1+\rho},{-2\over 1+\rho}\big)$
in powers of $\rho$, as is needed in \kisuy.
To simplify the result, we will use the parameter
$$
	b = c+2\,.
$$
We find the following 
expansions in $b$ of the functions\ $A(q,b-2), B(q,b-2)$\ \xcsdew\
involved in \fsre:
$$\eqalign{
&A(q,b-2) = {\Gamma(2-b-q)\, \Gamma(b)\over\Gamma(2-b)\Gamma(1+b-q)}\ 
\bigg\{ 1 + b^2\Big( {\pi^2\over6} + 
{ \psi(1-q)+\gamma_E\over q} \Big) + O(b^3) \bigg\}\, ,
\cr
&B(q,b-2) = {\Gamma(q+b)\Gamma(b)\over\Gamma(q)\Gamma(2b)}
\bigg\{ 1 +
{b\over2q} + {b^2\over2q^2}\big(q^2\psi'(q) - 1\big) 
+ {b^3\over4q}\big(q\psi''(q)+2\psi'(q)\big) + O(b^4) \bigg\} .
}$$
Hence,
$$
\eqalign{J_n(a,b-2) =&  {\pi^2\over2}\,{4a^2-n^2+2nb\over b^2(1-b)^2}
\exp\bigg\{ -G_n(a)\, b +\cr & \Big(\, {G_n''(a)\over12} -
 2\ {a\, G_n'(a)+G_n(a)\over n^2-4a^2} + 
{10\over3}\, \zeta(3) \Big)\, b^3 + O(b^4)\,  \bigg\}\ ,}
$$
where
$$
G_n(a) = 
\psi(a+n/2)+\psi(-a+n/2)+2\gamma_E \ ,$$
and $   G_n'(a) =
 {d\over da}\, G_n(a)\,   ,\ 
\  G_n''(a) = {d^2\over da^2}\, G_n(a)\, .
$

\appendix{B}{}

In this appendix, 
we write down explicit expressions for the coefficients
$u_1$, $u_2$, $v_1$ and $v_2$ taking part in
the expansion \krew, \vder\ of the 
the function $\Gamma_g$.

On the one hand, 
from the assumption \lskit, the coefficients of odd powers of $\rho$ in
the exponential factor of $Z_{n,\omega}$ \uytre\ are completely fixed
by the conjectured constant ${\bf Z}_n(\omega\beta)$ \lsidiiu.
This fixes $u_1,u_2$ uniquely, giving
\eqn\uu{\eqalign{
u_1 =& \Big(\omega^2-
{n^2\over16}\Big)\,  \Big(T_n(2\omega) 
-{3\over2}\Big) + {n(n-2)\over16}\, ,
\cr
u_2 =& 
{1\over12}\Big(\omega^2-{n^2\over16}\Big)
\Big(\omega^2+{n^2\over16}-{1\over2}\Big)\, T_n''(2\omega)
+ {\omega(4\omega^2-1)\over12}\, T_n'(2\omega)
+ {1\over4}\Big(\omega^2-{1\over12}\Big)T_n(2\omega)
\cr
& - {n(n+4)\over768} - {11\omega^2\over48}
+{1\over24}\, 
{\bar\tau}_0 + {\tau_2\over2}\, \Big(\omega^2-{n^2\over16}\Big)\, ,
}}
where
$$\eqalign{
&T_n(a) = 
\psi(a+n/2) + \psi(-a+n/2)+2\gamma_E 
+ 2\bar{\tau}_0 \,,\; 
\cr&T_n'(a) = {d\over da}\, T_n(a) \,,\;\ \ \ 
T_n''(a) = {d^2\over da^2}\,  T_n(a) \,.
}$$

On the other hand, the expansion in powers of $\rho$ of \kisuy\ 
uniquely determines 
the coefficients $u_1,\, v_1$ and 
$v_2-{3\over2}\, u_2$ in the first equation of \vder. The
coefficient $u_1$ thus 
obtained is in agreement with \uu, verifying the assumption \lskit\ 
to first order. 
The coefficients $v_1$ and (using the expressions $u_{1,2}$ from \uu) 
$v_2$ are
$$\eqalign{
v_1 &= {\omega\over2}\, \Big(\omega^2-{n^2\over16}\Big)\, T_n'(2\omega) +
{1\over4}\, \Big(\omega^2-{n^2\over16}\Big)\,  T_n^2(2\omega)
\cr & - {3\over4}\, \Big(\omega^2 - {n(5n-4)\over48}\Big)\, T_n(2\omega) +
{7\over8}\, \Big(\omega^2 - {n(17n-20)\over112}\Big)  + {u_1\over2}\ ,
\cr v_2 &= 
\Big(\omega^2-{n^2\over16}\Big)\, 
\Big({1\over24}-{\omega^2\over4}\Big)\, T_n''(2\omega) 
\cr& - \bigg( \Big(\omega^2-
{n^2\over16}\Big)\, {T_n(2\omega)\over2} +
 {1\over4}\, \Big(\omega^2+{n(n-4)\over16}-{1\over2}\Big)\bigg)\, 
\omega\,  T_n'(2\omega)
\cr & - {1\over12} \Big(\omega^2-{n^2\over16}\Big) T_n^3(2\omega) -
 {1\over8}\, \Big(\omega^2+{n(n-4)\over16}\Big)\, T_n^2(2\omega)-
 {1\over8}\, \Big(\omega^2-{n(n-2)\over8}-{1\over2} \Big)\, T_n(2\omega)
\cr &	 - {1\over8}\Big(\omega^2-{n^2\over16}\Big)\
\big( 2\, \tau_2-14\, \zeta(3)-3\big) -
 {n(n-8)\over256} + 
{u_1\over8} + {v_1\over2} + {3u_2\over2} - {{\bar\tau}_0\over8}\ .
}
$$

\appendix{C}{}

In this appendix, we give the formula for 
the three-particle contribution $F^{(3)}$
\bsgr\ to the fermion two-point function in the
$SU(2)$-Thirring model that we used for our numerical calculations.
We first  specialize
the expression written in \LZtopff\ to the case of  three-particle 
form factors of the field ${\cal O}_{\beta/ 4}^{1}$ for
$\beta^2=1$:
\eqn\jaasdy{\eqalign{&
\langle\, vac\,|\, {\cal O}_{1/4}^{ 1}(0)\, |\,
A_{-}(\theta_{1})\ldots A_{+}(\theta_{k})\ldots A_{-}(\theta_{3})\, 
\rangle_{in}=- {A_G^{9\over 2}\ \Gamma^{3}({1\over 4})\over
2^{15\over 4}\, e^{3\over 8}\, \pi^{9\over 4}}\ e^{{i\pi \over 8}}\
M^{1\over 4}\times\cr&
\prod_{m=1}^{3}
e^{ {\theta_m\over 4}}\ 
\prod_{m<j} G(\theta_m-\theta_j)\ 
\bigg\{\, 
 \int_{C_+}{d\gamma\over 2\pi} \, e^{-{\gamma\over 2}}
\prod_{p=1}^{k}W(\theta_p-\gamma)\prod_{p=k+1}^{3}
W(\gamma-\theta_p)+\cr &
\, \int_{C_-}{d\gamma\over 2\pi} \,
e^{-{\gamma\over 2}}\
\prod_{p=1}^{k-1}W(\theta_p-\gamma)\prod_{p=k}^{3}
W(\gamma-\theta_p) \, \bigg\}\ .}}
Here the functions $G$ and $W$ are 
\eqn\lsdyy{G(\theta)=
i\ { 2^{-{5\over 12}}\, e^{1\over 4}\, \Gamma({1\over 4})\over
\sqrt{\pi}\, A_G^{3}}\ \sinh(\theta/2)\ \exp\Big(\int_{0}^{\infty}
{dt\over t}\ {\sinh^2 t(1-i\theta/\pi)\, e^{-t}\over
\sinh(2t)\ \cosh(t)}\ \Big) }
and
\eqn\lsdy{
W(\theta)=2\ {\Gamma({3\over 4}-{i\theta\over 2\pi})\Gamma(-{1\over 4}+
{i\theta\over 2\pi})\over \Gamma^2({1\over 4})}\ .}
The contour $C_+$ starts from $-\infty$ 
on the real axis of the complex 
$\gamma$-plane, goes above the poles located
at $\gamma=\theta_p+i\pi/2$, 
$p=1,\ldots,k$ and below those located at $\gamma=\theta_p-i\pi/2$,
$p=k+1,\ldots, 3$,
always staying in the strip $-\pi/2-0 < \Im m\, \gamma < \pi/2+0$,
and finally extends to $+\infty$ on the real axis. 
Similarly, the contour $C_-$ goes above the poles
located at $\gamma=\theta_p+i\pi/2$, $p=1,\ldots,k-1$ 
and below those at $\gamma=\theta_p-i\pi/2$,\ 
$p=k,\ldots,3$. Notice that the 
integrals in\ \jaasdy\ can be expressed in terms
of the generalized hypergeometric function ${}_3F_2$ at unity.

Using the expressions\ \jaasdy\ and
performing one of the rapidity integrals in \dfre, one can
obtain the following form for
the function $F^{(3)}$ in\ \bsgr:
$$\eqalign{&
F^{(3)}=
{2\, e^{-{3\over 4}} A_G^{9}\over 3\pi\,  \Gamma^{6}({1\over 4})}
\int_{-\infty}^{\infty}\, 
    dxdy\ 
   { e^{ -Mr\sqrt{3+2\cosh x+2\cosh y + 2\cosh(x-y)} }
 \over (3+2\cosh x+2\cosh y + 2\cosh(x-y))^{1\over 4} }\times
\cr &
\big(\, 
2\, |R_1(x,y)|^2+|R_2(x,y)|^2\, \big)\ 
|G(x)G(y)G(x-y)|^2\, 
e^{x+y\over 2}\ 
\bigg({e^{-x}+e^{-y}+1\over e^x+e^y+1}\bigg)^{1\over 4}\ .}
$$
The functions $R_1$ and $R_2$ here are
$$
 R_2(x,y) = e^{-{x\over2}+{i\pi\over 4}}\ 
R_1(-x,y-x)-e^{-{y\over2}-{i\pi\over 4}\ }R^*_1(-y,x-y)
$$
and
$$\eqalign{
&  R_1(x,y) = 
- {\cosh{x\over2}\cosh{y\over2}\over 
2\sinh x\sinh y}\  
U\Big(-{1\over2},-{1\over2}-{ix\over2\pi},-{1\over2}-{iy\over2\pi};
-{ix\over2\pi},-{iy\over2\pi}\Big)
\cr&\ \   + e^{-{x\over 2}} 
{\cosh{y-x\over2}\cosh{x\over2}\over 2\sinh (y-x)\sinh x}\ 
U\Big({1\over2},{1\over 2}-{i(y-x)\over2\pi},{1\over2}+{ix\over 2\pi};
1-{i(y-x)\over 2\pi},2+{ix\over 2\pi}\Big)
\cr&\ \         + e^{-{y\over 2}} {\cosh{x-y\over2}\cosh{y\over2} 
\over 2\sinh (x-y)\sinh y}\ 
U\Big({1\over2},{1\over2}-{i(x-y)\over2\pi},
{1\over2}+{iy\over2\pi};1-{i(x-y)\over2\pi},2+{iy\over2\pi}\Big)\ ,
}$$
where\ 
$U(a,b,c;d,e)$ is related to the 
generalized hypergeometric function ${}_3F_2$ by
$$
U(a,b,c;d,e) =
{\Gamma(a)\Gamma(b)\Gamma(c)\over \Gamma(d)\Gamma(e)}\ 
{\ }_3F_2(a,b,c;d,e;1)\ .
$$

\listrefs

\end